# Cayley Configuration Spaces of 1-dof Tree-decomposable Linkages, Part II: Combinatorial Characterization of Complexity

Meera Sitharam, Menghan Wang, Heping Gao [1]

*University of Florida, Department of Computer Information Science and Engineering, Gainesville, Florida, USA*

**Abstract**

We continue to study *Cayley configuration spaces* of *1-degree-of-freedom (1-dof) linkages* in 2D begun in Part I of this paper, i.e. the set of attainable lengths for a non-edge. In Part II, we focus on the algebraic complexity of describing endpoints of the intervals in the set, i.e., the *Cayley complexity*.

Specifically, we focus on Cayley configuration spaces of a natural class of 1-dof linkages, called *1-dof tree-decomposable linkages* (defined in Part I, Section 2). The underlying graphs $G$ satisfy the following: for some *base non-edge* $f$, $G \cup f$ is *quadratically-radically solvable (QRS)*, meaning that $G \cup f$ is minimally rigid, and given lengths $\bar{l}$ of all edges, the corresponding linkage $(G \cup f, \bar{l})$ can be simply realized by ruler and compass starting from $f$. It is clear that the Cayley complexity only depends on the graph $G$ and possibly the non-edge $f$. Here we ask whether the Cayley complexity depends on the choice of a base non-edge $f$. We answer this question in the negative, thereby showing that low Cayley complexity is a property of the graph $G$ alone (independent of the non-edge $f$).

Then, we give a simple characterization of graphs with low Cayley complexity, leading to an efficient algorithmic characterization, i.e. an efficient algorithm for recognizing such graphs.

Next, we show a surprising result that (graph) planarity is equivalent to low Cayley complexity for a natural subclass of 1-dof tree-decomposable graphs. While this is a finite forbidden minor graph characterization of low Cayley complexity, we provide counterexamples showing impossibility of such finite forbidden minor characterizations when the above subclass is enlarged.

1. **Introduction**

We focus on *Cayley configuration spaces* (see Definitions in Part I, Section 1 and 2) of a natural class of 1-dof linkages, called *1-dof tree-decomposable linkages*, whose underlying graphs being *1-dof tree-decomposable graphs* (Fudos and Hoffmann, 1997; Owen and Power, 2007).

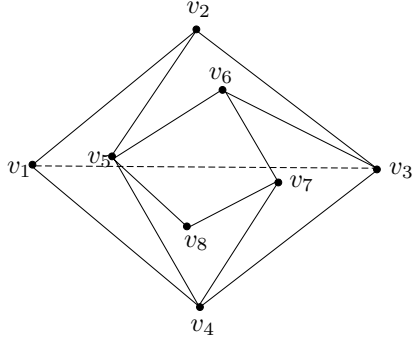

Figure 1. A 1-dof tree-decomposable graph with low Cayley complexity

In Part I, Section 1 we have defined various measures of complexity of a 1-dof graph $G$ in terms of Cayley configuration spaces over a non-edge $f$, that is, (a) Cayley size, the maximum number of intervals in the Cayley configuration space of any linkage with $G$ as underlying graph, (b) Cayley computational complexity, the maximum time complexity for obtaining all the interval endpoints of the Cayley configuration space of any linkage with $G$ as underlying graph, (c) Cayley complexity, the maximum algebraic complexity of describing each interval endpoint of the Cayley configuration space of any linkage with $G$ as underlying graph. In Part I of the paper, we have investigated Measures (a) and (b) for 1-dof tree-decomposable graphs. In Part II, we focus on Measure (c), Cayley complexity.

We urge readers to familiarize themselves with definitions of the following from Part I: linkages, rigidity and dofs (Part I, Section 1), 1-dof tree-decomposable graphs, base non-edges and construction steps (Part I, Section 2), extreme graphs (Part I, Section 3). Recall additionally from Part I that a 1-dof tree-decomposable graph $G$ has *low Cayley complexity* on non-edge $f$ (see example in Figure 1), if all of its *extreme graphs* occurring in the construction of $G$ from $f$ are tree-decomposable (Part I, Section 3).

Equipped with these concepts and definitions from Part I, Part II is self-contained and answers the following questions.

1.1. *Questions*

(1) Which 1-dof tree-decomposable graphs have low Cayley complexity? Can we find a finite forbidden minor characterization (Diestel, 2005) for such graphs?
    For this question to be well posed, we need to first settle the following question:
(2) Does Cayley complexity depend on the choice of non-edge $f$?
    As mentioned in Part I, Section 2, a 1-dof tree-decomposable graph $G$ can have multiple base non-edges. I.e., there could be non-edges $f \neq f'$ such that both $G \cup f$ and $G \cup f'$ are tree-decomposable graphs. For example, for the graph in Figure 1,

we can choose $(v_i, v_{i+2})$ as base non-edge for any $i$. Therefore we ask: can $G$ have different Cayley complexities on $f$ and $f'$?

## 1.2. Contributions

We first answer Question 2 in the negative in Section 3. Specifically, we show that if the extreme graphs are all tree-decomposable for some choice of $f$, then they are tree-decomposable for any non-edge $f$. This shows robustness of the Cayley complexity measure for 1-dof tree-decomposable graphs, and makes it sufficient to characterize low Cayley complexity of the graph with a specific convenient non-edge $f$. In addition, in Section 4 we give a simple characterization of 1-dof tree-decomposable graphs with low Cayley complexity yielding an efficient recognition algorithm.

To answer Question 1, we show a surprising result in Section 5 that (graph) planarity is equivalent to low Cayley complexity for a natural subclass of 1-dof tree-decomposable graphs. While this is a finite forbidden minor graph characterization of low Cayley complexity, we give counterexamples in Section 5.2 showing impossibility of such finite forbidden minor characterizations when the above subclass is enlarged.

## 1.3. Organization of Part II

In Section 2, we give additional, required definitions related to 1-dof tree-decomposable graphs, beyond those given in Part I, Sections 1, 2, 3.

In Section 3, we prove that the choice of base non-edge does not affect low Cayley complexity.

In Section 4, we give a characterization of 1-dof tree-decomposable graphs with low Cayley complexity, yielding an efficient recognition algorithm.

In Section 5, we study finite forbidden minor characterizations of low Cayley complexity. We introduce two natural subclasses: 1-path and triangle-free. In Section 5.1, we prove that low Cayley complexity is equivalent to planarity for 1-path, triangle-free, 1-dof tree-decomposable graphs. In Section 5.2, we show that finite forbidden minor characterization is impossible for more general classes of graphs of low Cayley complexity.

## 2. Definitions and basic properties of 1-dof tree-decomposable graphs

As a quick recall of the definitions from Part I, refer to Figure 2. The graph $G$ in Figure 2(a) is 1-dof tree-decomposable, with $f = (v_0, v'_0)$ as base non-edge. Two of the construction steps of $G$ are $v_1 \triangleleft (v_0 \in T_1, v'_0 \in T_2)$ and $v_7 \triangleleft (v_5, v_6)$. Recall that the $T_i$'s refer to maximal tree-decomposable subgraphs or *clusters* used in the construction of $G$, and $G_f(k)$ refers to the graph constructed from $f$ after the $k$th construction step. Thus, the above two steps yield graphs $G_f(1)$ and $G_f(7)$ respectively. We use $cdeg(v_7) = 3$ to denote that $v_7$ is shared by three clusters. Figure 2(b) is $G$'s extreme graph corresponding to Construction Step 7 from $f$, obtained by adding the edge $(v_5, v_6)$ to $G_f(6)$. This extreme graph is denoted $\hat{G}_f(7)$. The graph $G$ has low Cayley complexity on $f$ since all of its extreme graphs for $f$ are tree-decomposable.

Here we give an additional definition that will be used in our subsequent discussions.

**Definition 2.1.** The construction of a 1-dof tree-decomposable graph from a base non-edge $f = (v_0, v'_0)$ forms a partial order, giving rise to *levels*. Each level is a set of vertices, where

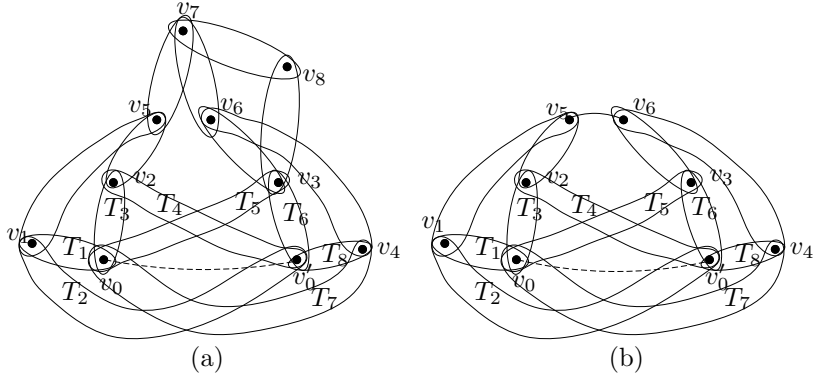

Figure 2. (a) A 1-dof tree-decomposable graph with base non-edge $(v_0, v_0')$ and levels $L_0$ to $L_4$. (b) Extreme graph $\hat{G}_f(7)$ for (a).

- Level 0 ($L_0$ in brief) consists of vertices $v_0$ and $v_0'$.
- Level 1 ($L_1$ in brief) is the set of vertices $v_i$ such that $v_i \triangleleft (v_0, v_0')$.
- Level $i$ ($L_i$ in brief) ($i \geq 2$) is the set of vertices $v_i$ such that $v_i \triangleleft (u, v)$, where $u \in L_j$, $w \in L_k$, the maximum of $j$ and $k$ is equal to $i - 1$.

The order of construction steps should be consistent with the partial order. Within the same level the construction steps can be arbitrarily ordered.

For example, refer to Figure 2(a). Vertices in $L_0$ are $v_0$ and $v_0'$). Vertices in $L_1$, namely those vertices constructed using only vertices in $L_0$, are $v_1 \triangleleft (v_0 \in T_1, v_0' \in T_2)$, $v_2 \triangleleft (v_0 \in T_3, v_0' \in T_4)$, $v_3 \triangleleft (v_0 \in T_5, v_0' \in T_6)$ and $v_4 \triangleleft (v_0 \in T_7, v_0' \in T_8)$. Similarly, vertices in $L_2$ are $v_5 \triangleleft (v_1, v_2)$ and $v_6 \in (v_3, v_4)$, in $L_3$ is $v_7 \triangleleft (v_5, v_6)$, and in $L_4$ is $v_8 \in (v_3, v_7)$.

We also recall from Part I the definition of the *last level*.

**Definition 4.4.***(Part I)* A vertex $v$ is in a 1-dof tree-decomposable graph $G$'s *last level* $L_t$ if: (a) $cdeg(v) = 2$, i.e. there are exactly two clusters $T_1$ and $T_2$ sharing $v$; (b) each of $T_1$ and $T_2$ has only one shared vertex with the rest of the graph $G' = G \setminus (T_1 \cup T_2) = G \setminus \{v\}$.

## 3. Characterizing parameter choices: all base non-edges yield Cayley configuration spaces of the same complexity

We know that a 1-dof tree-decomposable graph $G$ can have multiple possible base non-edges, as seen in Figure 1. Can $G$ have different Cayley complexities on different base non-edges? The following theorem shows that this is impossible.

**Theorem 3.1.** *A 1-dof tree-decomposable graph $G$ either has low Cayley complexity on all possible base non-edges, or on none of them.*

**Idea of the proof.** We prove by contradiction, assuming that there exists a minimum graph $G$ with low Cayley complexity on one base non-edge, but not on another base non-edge. We then show the minimality cannot hold, since we can obtain a smaller graph with the same property by leaving out a final construction step from $G$.

**Proof.** Assume that we can pick a graph $G$ with the minimum number of construction steps, such that $G$ has two base non-edges $f = (v_1, v_2)$ and $f' = (v'_1, v'_2)$, with low Cayley complexity on $f$ but not on $f'$.

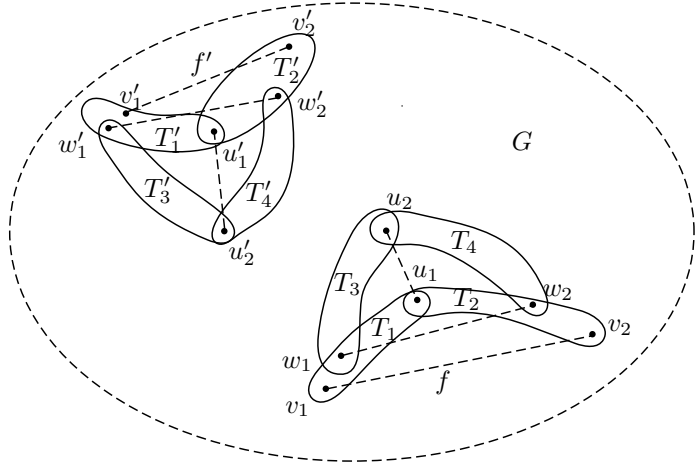

Figure 3. For proof of Theorem 3.1

The two base non-edges $f$ and $f'$ cannot connect the same pair of clusters, otherwise $G$ will have the same Cayley complexity on them. Therefore, $G$ must have more than one construction step, whether from $f$ or $f'$. Let the first two construction steps from $f$ be $u_1 \triangleleft (v_1 \in T_1, v_2 \in T_2)$ and $u_2 \triangleleft (w_1 \in T_3, w_2 \in T_4)$, the first two construction steps from $f'$ be $u'_1 \triangleleft (v'_1 \in T'_1, v'_2 \in T'_2)$ and $u'_2 \triangleleft (w'_1 \in T'_3, w'_2 \in T'_4)$. See Figure 3. Notice that $(v_1, v_2)$ and $(w_1, w_2)$ may or may not be the same pair of vertices, but they must yield the same Cayley complexity for $G$ since they connect the same pair of clusters, so do $(v'_1, v'_2)$ and $(w'_1, w'_2)$. So without loss of generality, we can choose $f = (w_1, w_2)$, $f' = (w'_1, w'_2)$.

Recall the definition of "*last level*" of vertices of a 1-dof tree-decomposable graph from Part I. We make the following claim:

**Claim 3.2.** *Let $v_k$ be in the last level $L_t$ of $G$, then $v_k$ must be a vertex of $f$ or $f'$.*

**Proof.** To the contrary, suppose there exists a last level vertex $v_k \triangleleft (w_m \in T_m, w_n \in T_n)$ such that $v_k$ is not a vertex of $f$ or $f'$.

Note that for any base non-edge presented in the graph $G' = G \setminus \{v_k\}$, $v_k \triangleleft (w_m \in T_m, w_n \in T_n)$ can be regarded as the last step for construction of $G$ from that base non-edge, and $G'$ can be viewed as the graph obtained from $G$ by leaving out $v_k \triangleleft (w_m, w_n)$. Thus the graph $G' = G_f(k-1) = G_{f'}(k-1)$ is a smaller 1-dof tree-decomposable graph with $f$ and $f'$ as two base non-edges, and the extreme graph of $G$ corresponding to $v_k$ is $\hat{G}_f(k) = \hat{G}_{f'}(k) = G' \cup (w_m, w_n)$.

Let $S$ be the set of all extreme graphs of $G$ for $f'$, and let $S_{k-1}$ be the set of all extreme graphs of $G'$ for $f'$. Now $S = S_{k-1} \cup \{\hat{G}_f(k)\}$. Since $G$ does not have low Cayley complexity on $f'$, at least one extreme graph in $S$ is not tree-decomposable; since $G$ has low Cayley complexity on $f$, $\hat{G}_f(k)$ is tree-decomposable, so at least one extreme graph in $S_{k-1}$ is not tree-decomposable. Thus $G'$ has low Cayley complexity on $f$, but does not have low Cayley complexity on $f'$, violating the minimality of $G$. Therefore no such $v_k$ exists. □

Let $v$ be the vertex of $G$ constructed last in a construction starting from $f = (w_1, w_2)$. By Claim 3.2, $v \in \{w_1', w_2'\}$. Without loss of generality, assume that $v = w_1'$.

Notice that $G$ has the same Cayley complexity on $f' = (w_1', w_2')$ and $f'' = (u_1', u_2')$, since from $(u_1', u_2')$ we can construct $w_1'$ and $w_2'$ first, and the subsequent constructions are all the same. Since $w_1'$ is not a vertex of $f$ or $f''$, using the argument of Claim 3.2, we can prove that the 1-dof tree-decomposable graph $G \setminus \{w_1'\}$ has low Cayley complexity on $f$, but not on $f''$, violating the minimality of $G$. Therefore, we conclude that no such minimal graph $G$ can exits, which has low Cayley complexity on one base non-edge and not on another base non-edge, thus proving Theorem 3.1. □

PSfrag replacements

The theorem shows that our measure of low Cayley complexity is robust: it depends only on the graph, not on the chosen non-edge. I.e., if we can characterize low Cayley complexity for a 1-dof tree-decomposable $G$ for a convenient base non-edge $f$, it translates to a characterization of low Cayley complexity of $G$ (for all possible base non-edges).

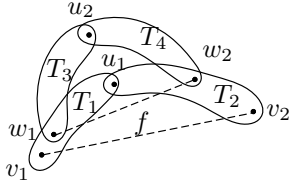

Figure 4. The first two construction steps starting from $f$: $(w_1, w_2)$ can be used as base non-edge instead. See Note.

**Note.** In the following discussion, when considering Cayley complexity for graphs with two or more construction steps, for simplicity we always pick a base non-edge $f$ with both endpoints being shared vertices, so that $L_1$ contains at least two construction steps. If the given base non-edge's endpoints $v_1$ and $v_2$ are not both shared vertices, and the first two construction steps are $u_1 \triangleleft (v_1 \in T_1, v_2 \in T_2)$ and $u_2 \triangleleft (w_1 \in T_1', w_2 \in T_2')$, as in Figure 4, we can always choose $(w_1, w_2)$ — the base pair of vertices for the second construction step — as base non-edge instead.

## 4. A characterization of 1-dof tree-decomposable graphs with low Cayley complexity and an efficient recognition algorithm

In this section we introduce Theorem 4.1, the "four-cycle theorem". It characterizes 1-dof tree-decomposable graphs with low Cayley complexity by the structure of construction, that is, a graph $G$ is 1-dof tree-decomposable if and only if there exists a "base four-cycle" for each construction step, as shown in Figure 5 (a).

**Theorem 4.1.** [four-cycle theorem] *A 1-dof tree-decomposable graph $G$ with six or more clusters has low Cayley complexity on a base non-edge $f$, if and only if every construction step $v_k \triangleleft (u_k, w_k)$ from $f$, where $v_k$ is in $L_2$ or higher levels, has its base pair of vertices* taken from an adjacent pair of clusters in a four-cycle of clusters. I.e., we can find four clusters $T_1, T_2, T_3, T_4$ in $G_f(k-1)$, such that $u \in T_1, w \in T_2$, $T_1 \cap T_2 = \{p_1\}$, $T_2 \cap T_3 = \{p_3\}$, $T_3 \cap T_4 = \{p_2\}$, $T_4 \cap T_1 = \{p_4\}$, where $p_1, p_2, p_3, p_4$ are distinct vertices. See Figure 5 (a). Note that a graph with less that six clusters trivially has low Cayley complexity.

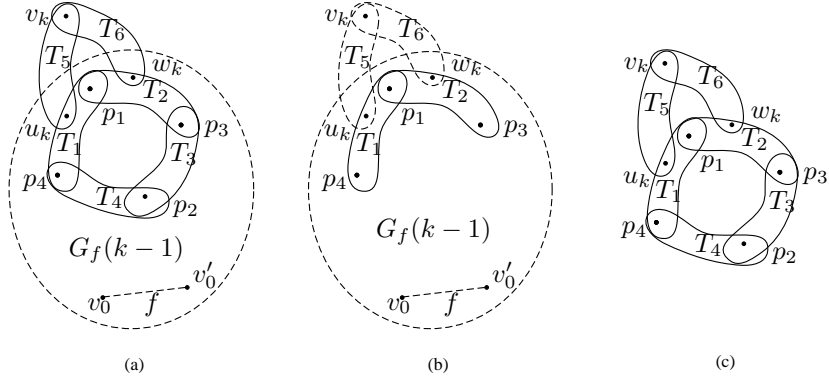

Figure 5. For theorem 4.1 (four-cycle theorem)

**Idea of the proof.** The $\Longrightarrow$ direction directly follows from the structure of extreme graphs, which are all tree-decomposable by definition of low Cayley complexity. For the $\Longleftarrow$ direction, we prove by induction on the number of construction steps, showing a tree-decomposable construction sequence for each extreme graph via the corresponding base four-cycle.

**Proof.**
($\Longrightarrow$): for a graph $G$ with low Cayley complexity, consider any vertex $v_k \triangleleft (u_k, w_k)$ in $L_2$ or higher levels. By the definition of low Cayley complexity, the extreme graph $\hat{G}_f(k)$ is tree-decomposable, so there exists a construction sequence of $\hat{G}_f(k)$ from $e_k = (u_k, w_k)$. Since $v_k \triangleleft (u_k, w_k)$ is in $L_2$ or higher levels, $G_f(k-1)$ has at least four clusters, thus $\hat{G}_f(k)$ has at least two construction steps. Refer to Figure 5 (b), let the first construction step of $\hat{G}_f(k)$ from $e_k$ be $p_1 \triangleleft (u_k \in T_1, w_k \in T_2)$. The next construction step must be constructed on cluster pairs $(T_1, T_2)$, yielding a four-cycle just as in Figure 5(a).
($\Longleftarrow$): we prove by induction on the number of construction steps. The base case, where $G$ has exactly six clusters, is shown in Figure 5 (c), which clearly has low Cayley complexity.

As induction hypothesis, assume the theorem holds for all tree-decomposable graphs with less than $k$ construction steps. For the induction step, consider a graph $G = G_f(k)$ with $k$ construction steps where every construction step is based on a four-cycle. By induction hypothesis, all of its extreme graphs $\hat{G}_f(i)$ with $i < k$ are tree-decomposable, so we only need to prove that $\hat{G}_f(k)$ is tree-decomposable. Let step $v_k$ be based on the four-cycle $T_1, T_2, T_3, T_4$ as in Figure 5 (a).

One pair of adjacent clusters in any cycle of clusters must appear (for the first time) in a single construction step from $f$. Otherwise, assume that each cluster in the cycle appears in separate construction steps, and call the cluster from the last of those steps $T_l$.

Notice that at the time of appearing in construction step, every cluster should have only one shared vertex with the previously constructed graph, which includes the remaining vertices of the cycle. But this is impossible for $T_l$ since it participates in the cycle.

Suppose the cluster pair $(T_1, T_2)$ appear together in a single construction step, the $m$th construction step $p_1 \triangleleft (u_m \in T_2, w_m \in T_1)$, where $m < k$. By induction hypothesis, the corresponding extreme graph $\hat{G}_f(m)$ is tree-decomposable, so from non-edge $(u_m, w_m)$ we can construct $G_f(m-1) = \hat{G}_f(m) \backslash (u_m, w_m)$, and then $G_f(k-1)$ by appending the $m$th to $(k-1)$th construction steps to $G_f(m-1)$. Since the entire four-cycle $T_1 T_2 T_3 T_4$, containing the non-edge $(u_m, w_m)$, can be constructed from $(u_k, w_k)$, we can also construct $G_f(k-1)$, thus $\hat{G}_f(k)$, from $(u_k, w_k)$. So $\hat{G}_f(k)$ is also tree-decomposable and $G_f(k)$ has low Cayley complexity.

For the other cases, where the cluster pair appearing together in a single construction steps is $(T_2, T_3)$, $(T_3, T_4)$ or $(T_4, T_1)$, the argument is similar. □

Theorem 4.1 yields an algorithm to verify whether a given 1-dof tree-decomposable graph $G$ has low Cayley complexity on base non-edge $f$, with expected time complexity linear in $|V|$. This is more efficient than the algorithm that follows the definition of low Cayley complexity, i.e., checking if all extreme graphs are tree-decomposable, which takes $O(|V|^3)$ time (checking $O(|V|)$ extreme graphs, each taking $O(|V|^2)$ time (Fudos and Hoffmann, 1997) ). The algorithm follows the construction of $G$ from $f$ and maintains a list $L$ of adjacent cluster pairs that can be used as base pair of clusters for the next construction step.

**Algorithm 1. Recognizing low Cayley complexity**
(1) For each construction step $v_k \triangleleft (u_k \in T_k, w_k \in T'_k)$ where $v_k \in L_1$, add $(T_k, T'_k)$ to $L$.
(2) For each subsequent construction step $v_k \triangleleft (u_k \in T_k, w_k \in T'_k)$, do:
   (1) Find the following two sets of clusters: $U = \{T_u | T_u \in G_f(k-1), u_k \in T_u\}$, $W = \{T_w | T_w \in G_f(k-1), w_k \in T_w\}$.
   (2) Find all the pairs $(T_u, T_w) \in U \times W$ such that $(T_u, T_w)$ is in $L$; if no such pair exists, return that $G$ does not have low Cayley complexity.
   (3) For all the pairs $(T_u, T_w) \in U \times W$ such that $T_u \cap T_w \neq \emptyset$, add $(T_k, T_u)$ and $(T'_k, T_w)$ to $L$.
   (4) Add $(T_k, T'_k)$ to $L$.

Let $n$ be the total number of construction steps. Then the total number of clusters is $2n$ since each construction step adds two clusters, and the total number of shared vertices is at most $3n$. Suppose we maintain a $2n \times 2n$ array storing the shared vertices of any two clusters, an array of length $3n$ storing the lists of clusters sharing each vertex, and represent $L$ as a hash-table. Step (1) takes $O(n)$ time. For (2), at each construction step $k$, step 1 takes $O(d)$ time, step 2 takes $O(d^2)$ time, step 3 takes $O(d^2)$ time, step 4 takes $O(1)$ time, where $d$ is the *cdeg* of $u_k$ and $w_k$. Therefore the overall time complexity is $O(n) + O(nd^2)$. Since the total number of clusters is $2n$, the average $d$ is constant, and the expected time complexity is linear in $n$, namely $O(|V|)$.

## 5. Finite forbidden minor characterization of low Cayley complexity

To go beyond algorithmic characterization of low Cayley complexity, we ask if there exists a *finite forbidden minor characterization* for 1-dof tree-decomposable graphs with low Cayley complexity.

**Definition 5.1** (Diestel 2005). A graph $G$ has a graph $K$ as *minor* if $G$ can be reduced to $K$ via vertex/edge deletions and edge contractions (coalescing or identifying the two vertices of an edge). A class $C$ of graphs $G$ has a *finite forbidden minor characterization*, if there exists a fixed, finite set $M$ of minors, such that $G \in C$ if and only if $G$ doesn't contain any $K \in M$ as a minor.

Unfortunately, there is no finite forbidden minor characterization for general 1-dof tree-decomposable graphs, as we will show in Section 5.2. So in order to obtain a finite forbidden minor characterization, we must limit ourselves to more specialized subclasses. We consider two such natural subclasses: *1-path* and *triangle-free*.

*1-path:* Recall that 1-path, 1-dof tree-decomposable graphs are used as the base case in the proof of Theorem 4.3 in Part I. An intuitive interpretation for 1-path 1-dof tree-decomposable graph is that for the given base non-edge, there is only one "direction" of construction. In other words, there is only one "path" towards the unique "last step". As an example, refer to Figure 6. Both (a) and (c) are 1-dof tree-decomposable with $f$ as base non-edge. (a) is 1-path, but (c) is not, since in (a) $v_3$ must be constructed before $v_4$, while in (c) $v_3$ and $v_4$ lie in two independent "paths" and can be constructed in an arbitrary order.

Next we repeat the formal definition of 1-path graphs.

**Definition 4.5.** *(Part I)* A 1-dof tree-decomposable graph $G$ has a *1-path construction* from base non-edge $f = (v_0, v_0')$ if there is only one shared vertex $v$ in the last level $L_t$, other than $v_0$ and $v_0'$. As long as there exists a base non-edge permitting 1-path construction, we say the 1-dof tree-decomposable graph $G$ has the *1-path* property.

As an example, refer to Figure 6. Choose base non-edge $f = (v_0, v_0')$. Both (a) and (b) are 1-path with $v_4$ as the only shared vertex in the last level $L_t$ other than $v_0$ and $v_0'$. Both (c) and (d) are not 1-path, since both $v_3$ and $v_4$ are in $L_t$. Both (a) and (c) have low Cayley complexity on $f$, while (b) and (d) do not, since both of them have the extreme graph $\hat{G}_f(4)$ not tree-decomposable.

**Note.** A 1-dof tree-decomposable graph could have 1-path construction from some base non-edges but not from others. An example is provided in Figure 7. If we choose $f_1$ as base non-edge, the graph construction is 1-path and $v_5 \triangleleft (v_3, v_4)$ must be the last construction step. But if we choose $f_2$ as base non-edge, the graph construction is no longer 1-path, since $v_1$, $v_2$ and $v_5$ can be constructed in an arbitrary order. For a graph to have the 1-path property, it is sufficient if there exists one base non-edge permitting a 1-path construction, since choosing the appropriate base non-edge does not affect low Cayley complexity (by Theorem 3.1).

*Triangle-free:* Notice that every tree-decomposable graph, with more vertices than a single edge, contains at least one triangle. Therefore, the clusters of a triangle-free 1-dof tree-decomposable graph are exactly its edges. See Figure 8.

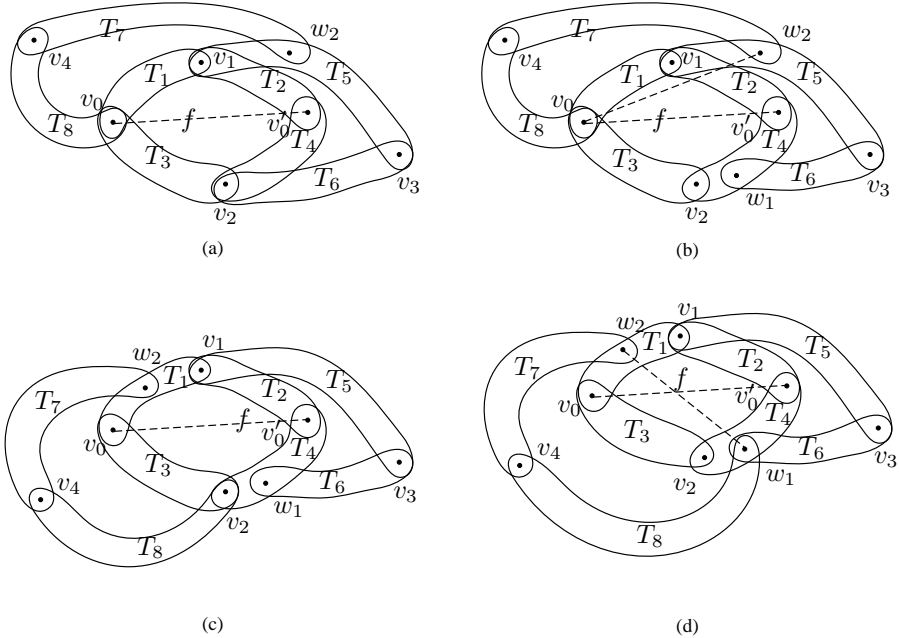

Figure 6. (a)(b) are 1-path and (c)(d) are multi-path 1-dof tree-decomposable graphs. (a)(c) have low Cayley complexity on $f$ while (b)(d) do not.

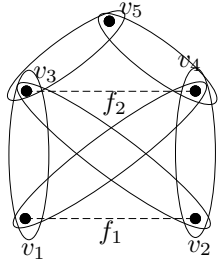

Figure 7. Choice of base non-edge does affect 1-path construction

**Remark.** A triangle-free 1-dof tree-decomposable graph is in fact equivalent to a *simple 1-dof Henneberg-I graph* (Gao and Sitharam, 2009). *Henneberg-I graphs* are graphs that can be constructed with a series of *Henneberg-I steps* from a base edge $f$. Each Henneberg-I step adds a new vertex with edges between it and exactly 2 previously constructed vertices. A *simple 1-dof Henneberg I graph* is obtained by removing a base edge from a Henneberg-I graph.

Next we give the formal definition for triangle-free graphs.

**Definition 5.2.** A 1-dof tree-decomposable graph $G$ is *triangle-free* if no subgraph of $G$ is a triangle.

Counterexamples in Section 5.2 show that finite forbidden minor characterization of low Cayley complexity does not exist for 1-path 1-dof tree-decomposable graphs, or triangle-free 1-dof tree-decomposable graphs.

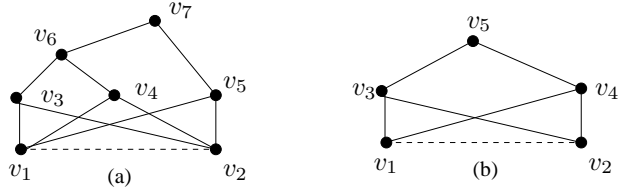

Figure 8. Both (a) (b) are triangle-free 1-dof tree-decomposable graphs. (b) has low Cayley complexity on $(v_1, v_2)$ while (a) does not.

On the other hand, for 1-dof tree-decomposable graphs that are both 1-path and triangle-free, we will show in the next section that surprisingly, planarity is equivalent to low Cayley complexity.

5.1. *Equivalence of low Cayley complexity to planarity for 1-path, triangle-free, 1-dof tree-decomposable graphs*

We prove the following theorem:

**Theorem 5.3.** *Let $G$ be a 1-path, triangle-free, 1-dof tree-decomposable graph with base non-edge $f = (v_0, v_0')$. $G$ has a configuration space of low Cayley complexity if and only if $G$ is planar.*

For the proof we use the familiar forbidden minor characterization of planar graphs by Kuratowski:

**Theorem 5.4** (Kuratowski 1930). *A graph $G$ is planar if and only if it contains no $K_{3,3}$ or $K_5$ minor.*

**Idea of the proof of Theorem 5.3.** We first divide all the 1-path, triangle-free, 1-dof tree-decomposable graphs into cases, according to the number of vertices in the first level $L_1$. We analyze these cases using Lemma 5.7 and 5.8, rule out the trivial cases and focus on a single case.

For the $\Longleftarrow$ direction, we actually prove a stronger version that $G$ has low Cayley complexity if $G$ contains no $K_{3,3}$ minor. We take the contrapositive and prove by contradiction, assuming that there exists a minimum graph $G$ without any $K_{3,3}$ minor not having low Cayley complexity, and then showing the minimality cannot hold.

For the $\Longrightarrow$ direction, we prove separately for $K_{3,3}$ and $K_5$. For $K_{3,3}$, we prove by contradiction, assuming that there exists a minimum graph $G$ with a $K_{3,3}$ minor having low Cayley complexity. We then analyze the possible vertex/edge deletions and edge contractions on $G$, showing that we cannot obtain a $K_{3,3}$ minor without violating the minimality of $G$, using Lemma 5.7 and 5.8. For $K_5$, the proof is similar to that of $K_{3,3}$.

**Proof. (of Theorem 5.3)**

Let $m$ denote the number of $G$'s vertices in $L_1$. We have four cases, of which one is trivial and two directly follow from Lemmas to be proved later.

[**Case 1**] $m \leq 1$, $G$ has at most three vertices. So $G$ trivially has low Cayley complexity on $f$ and it has neither minor, hence the theorem holds.

[**Case 2**] $m \geq 3$, by Lemma 5.7 (1.a) and Lemma 5.7 (2.a), $G$ has a $K_{3,3}$ minor and $G$ does not have low Cayley complexity on $f$, hence the theorem holds.

[**Case 3**] $m = 2$, $deg(v_0) \geq 3$, $deg(v_0') \geq 3$, by Lemma 5.7 (1.b) and Lemma 5.7 (2.b), $G$ has a $K_{3,3}$ minor and $G$ does not have low Cayley complexity on $f$, hence the theorem holds.

Therefore in the remainder of the proof, we assume that $G$ is in Case 4, as below.

[**Case 4**] $m = 2$, and, without loss of generality, either **(a)** $deg(v_0) = 2$, $deg(v_0') > 2$, or **(b)** $deg(v_0) = 2$, $deg(v_0') = 2$.

Let the two $L_1$ vertices be $v_1, v_2$. By Lemma 5.8, for (a), $G' = G \setminus \{v_0\}$ is a 1-path, triangle-free, 1-dof tree-decomposable graph with base non-edge $(v_1, v_2)$; for (b), $G' = G \setminus \{v_0, v_0'\}$ is a 1-path, triangle-free, 1-dof tree-decomposable graph with base non-edge $(v_1, v_2)$. In either case, due to Theorem 3.1, $G$ has low Cayley complexity on $(v_0, v_0')$ if and only if $G'$ has low Cayley complexity on $(v_1, v_2)$.

($\Longleftarrow$): we prove the contrapositive that if $G$ does not have low Cayley complexity on $(v_0, v_0')$, $G$ must have a $K_{3,3}$ minor, thus not planar. We prove by contradiction. To the contrary, assume that we can choose a $G$ with minimum number of vertices such that $G$ does not have low Cayley complexity on $(v_0, v_0')$ and $G$ does not have a $K_{3,3}$ minor. Since $G$ does not have low Cayley complexity on $(v_0, v_0')$, $G'$ does not have low Cayley complexity on $(v_1, v_2)$. Graph $G$ has no $K_{3,3}$ minor, so neither does $G'$. Therefore $G'$ has less number of vertices than $G$, does not have low Cayley complexity on the base non-edge and does not have a $K_{3,3}$ minor, contradicting the minimality of $G$.

($\Longrightarrow$): the proof splits into two sections.

*Proof that low Cayley complexity implies no $K_{3,3}$ minor*: again we prove by contradiction. Assume to the contrary that we can choose a $G$ with minimum number of vertices such that $G$ has low Cayley complexity on $(v_0, v_0')$ and $G$ has a $K_{3,3}$ minor.

Case 4(a): refer to Figure 9(a). For $G$ to be triangle-free, the third step vertex $v_3$ must be constructed on $(v_1, v_2)$. Recall that the $K_{3,3}$ minor is obtained from $G$ by a series of vertex/edge deletions and edge contractions. First notice that we cannot delete $v_0$ in this process, otherwise the result will be a minor of $G'$, which means that $G'$ also contains a $K_{3,3}$, contradicting the minimality of $G$. Since $deg(v_0) = 2$ and $K_{3,3}$ has no degree 2 vertices, either $(v_0, v_1)$ or $(v_0, v_2)$ must be contracted, and we obtain a graph $G_2$ as in Figure 9(b).

PSfrag replacements

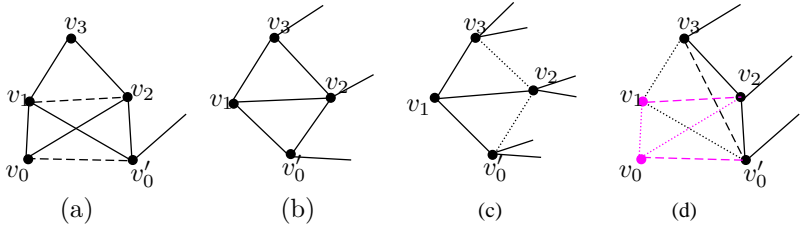

Figure 9. For proof of Theorem 5.3. (b) $G_2$: $(v_0, v_1)$ (or $(v_0, v_2)$) is contracted. (c) $G_3$: $(v_2, v_3)$ and $(v_2, v_0')$ are deleted. (d) $G'' = G' \setminus \{v_1\}$.

The two triangles in $G_2$: $\triangle(v_1, v_2, v_0')$ and $\triangle(v_1, v_2, v_3)$ cannot appear in $K_{3,3}$, so some vertex/edge deletion or edge contraction must eliminate them in order to end up in the $K_{3,3}$ minor. Notice that the edge $(v_1, v_2)$ must be preserved in this process, otherwise the result will again be a minor of $G'$, contradicting the minimality of $G$. Observe also that: (1) Deleting any of the four vertices $\{v_1, v_2, v_0', v_3\}$ will produce a subgraph of $G'$, contradicting the minimality of $G$. (2) Contracting any of the five edges of the two triangles will identify $(v_1, v_2)$ with another edge and produce a minor of $G'$, contradicting the minimality of $G$. Consequently, we must delete one edge from each triangle (excluding $(v_1, v_2)$).

Now notice since $G'$ is a 1-path triangle-free 1-dof tree-decomposable graph with low Cayley complexity on $(v_1, v_2)$, by Lemma 5.7, one of $v_1, v_2$, without loss of generality say $v_1$, must have degree 2 in $G'$, so $deg(v_1) = 3$ in both $G$ and $G_2$. We now go back to deleting edges from the two triangles in $G_2$. If we delete any edge adjacent to $v_1$, $deg(v_1)$ will become 2, in that case we must contract $v_1$ and the result will be a minor of $G'$, contradicting minimality of $G$. The only remaining way to obtain a $K_{3,3}$ minor is deleting both $(v_2, v_3)$ and $(v_2, v_0')$. Call the resulting graph $G_3$. We now give the proofs of Lemma 5.7.

Let $G'' = G' \setminus \{v_1\}$. See Figure 9(d). Notice that $v_3$ and $v_0'$ have the same degree in $G''$ and $G_3$. Since $deg(v_1) = 2$ in $G'$, by Lemma 5.8, $G''$ is a 1-path 1-dof triangle-free tree-decomposable graph with low Cayley complexity on $(v_0', v_3)$. By Lemma 5.7, at least one of $\{v_0', v_3\}$ must have degree 2 in $G''$, thus also in $G_3$. However notice that if any of $\{v_2, v_0', v_3\}$ has degree smaller than 3, $G_3$ would be a minor of $G'$, contradicting the minimality of $G$. Therefore $G$ cannot contain a $K_{3,3}$ minor.

Case 4(b): the argument is similar to the above.

*Proof that low Cayley complexity implies no $K_5$ minor:* we prove by contradiction. Assume to the contrary that we can choose a $G$ with minimum number of vertices such that $G$ has low Cayley complexity on $(v_0, v_0')$ and $G$ has a $K_5$ minor.

Case 4(a): as the proof above of no $K_{3,3}$ minor, since $K_5$ has no degree 2 vertex, $v_1$ must be contracted, and $deg(v_1) = 3$ in the contracted graph $G_2$ as in Figure 9 (b). However, since every vertex in $K_5$ has degree 4, $v_1$ must be either deleted or contracted. In either case, we will obtain a minor of $G'$. Thus $G'$ also contains a $K_5$ minor, contradicting the minimality of $G$.

Case 4(b): the argument is similar to the above.

Therefore, if $G$ has low Cayley complexity, $G$ can contain neither $K_5$ nor $K_{3,3}$, thus is planar. So we have proved Theorem 5.3. □

We now give the proofs of Lemmas 5.7 and 5.8 used in the above proof. These in turn depend on two other Lemmas 5.5 and 5.6.

**Lemma 5.5.** *Given a 1-dof tree-decomposable graph $G$ with base non-edge $f = (v_0, v_0')$, no minimally rigid subgraph $G'$ of $G$ can contain both $v_0$ and $v_0'$.*

**Proof.** For the sake of contradiction, assume there exists a 1-dof tree-decomposable graph $G$ with base non-edge $f = (v_0, v_0')$ such that both $v_0$ and $v_0'$ are contained in a minimally rigid subgraph $G'$ of $G$.

A graph $G = (V, E)$ in 2D is *minimally rigid*, if it satisfies the Laman condition (Laman, 1970), i.e., $|E| = 2|V| - 3$ and $|E_s| \leq 2|V_s| - 3$, for all subgraphs $G_s = (V_s, E_s)$ of $G$. Since $G'$ is minimally rigid, the graph $G' \cup f = (V_s, E_s)$ will have $|E_s| > 2|V_s| - 3$, contradicting the fact that $G' \cup f$ is a subgraph of the minimally rigid graph $G \cup f$. □

**Lemma 5.6.** *Let $G$ be a 1-path, 1-dof tree-decomposable graph with base non-edge $f = (v_0, v'_0)$. Then $G$ does not have low Cayley complexity on $f$ if either of the following holds:*
  (1) *The number of vertices in level $L_1$ is 3 or more.*
  (2) *The number of vertices in level $L_1$ is exactly 2, and both $v_0$ and $v'_0$ are not in the last level $L_t$.*

**Proof.** Let $v_n \triangleleft (u, w)$ be the last construction step from $f$. Let $m$ be the number of vertices in $L_1$. Since $G$ is 1-path, $v_n$ cannot be in $L_1$ when $m > 1$.

**1.** We prove by contradiction, showing that the extreme graph $\hat{G}_f(n)$ of $G$ is not tree-decomposable.

Assume that $\hat{G}_f(n)$ has a triangle decomposition into components $C_1$, $C_2$ and $C_3$. First, $v_0$ and $v'_0$ cannot lie in the same component for the following reason: for the sake of contradiction, assume without loss of generality that both $v_0$ and $v'_0$ are in $C_1$. By Lemma 5.5, $C_1$ is not minimally rigid unless it is not a subgraph of $G$, i.e., $C_1$ must contain the extreme edge $e_n = (u, w)$. Let $C_2 \cap C_3 = \{v_1\}$. Refer to Figure 10(a), in $G$ we have $v_1 \in L_t$ and $v_1 \notin \{v_n, v_0, v'_0\}$, contradicting the condition that $G$ is 1-path.

So without loss of generality assume that $v_0 \in C_1$, $v'_0 \in C_2$. Observe that $v_0$ and $v'_0$ must both be shared vertices between components or they will lie in the same component. Consider construction steps $v_i \triangleleft (v_0 \in T_i, v'_0 \in T'_i)$, where $v_i \in L_1$. Notice that no cluster $T_i$ or $T'_i$ can straddle two or more components, otherwise it will not be minimally rigid. Therefore as illustrated in Figure 10(b), we can have at most two $L_1$ vertices, contradicting the condition that $m \geq 3$.

**2.** We again prove by contradiction. Assume $\hat{G}_f(n)$ has a triangle decomposition into $C_1$, $C_2$ and $C_3$. As shown in the proof of (1), $v_0$, $v'_0$ cannot lie in the same component. In order to have two $L_1$ vertices, the first two vertices constructed, $v_1$ and $v_2$, have to be shared vertices between the components. One vertex of the base non-edge, without loss of generality say $v'_0$, is in $C_2$ as a non-shared vertex and the other, $v_0$, is the shared vertex between $C_1$ and $C_3$, as in Figure 10(b).

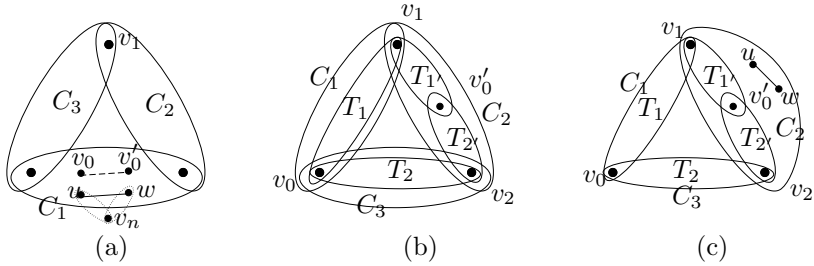

Figure 10. For proof of Lemma 5.6 (2).

Observe that $G$ can also take $(v_1, v_2)$ as base non-edge. Since $C_2$ contains a base non-edge $(v_1, v_2)$, by Lemma 5.5, $C_2$ must contain the extreme edge $e_n = (u, w)$ to be minimally rigid. This implies that tree-decomposable components $C_2$ and $C_3$, not containing $e_n$, are tree-decomposable subgraphs of $G$. So $T_1 = C_1, T_2 = C_3$ (see Figure 10)(c). We can see that $v_0$ is in the last level $L_t$ of $G$, contradicting the condition that both $v_0$ and $v_0'$ are not in $L_t$. □

**Lemma 5.7.** *(1) Let $G$ be a 1-path, triangle-free, 1-dof tree-decomposable graph with base non-edge $f = (v_0, v_0')$. Then $G$ has a $K_{3,3}$ minor if either of the following holds:*

*1.a the number of vertices in level $L_1$ is 3 or more.*

*1.b the number of vertices in level $L_1$ is exactly 2 and $deg(v_0) \geq 3$, $deg(v_0') \geq 3$.*

*(2) Let $G$ be a 1-path, triangle-free, 1-dof tree-decomposable graph with base non-edge $f$. Then $G$ does not have low Cayley complexity on $f$ if either of the following holds:*

*2.a the number of vertices in level $L_1$ is 3 or more.*

*2.b the number of vertices in level $L_1$ is exactly 2 and $deg(v_0) \geq 3$, $deg(v_0') \geq 3$.*

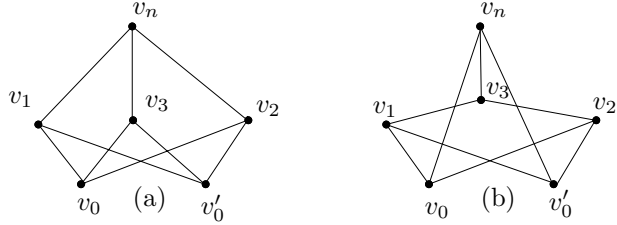

Figure 11. Edge contractions for Lemma 5.7.

**Proof.**
**1.a** Let $v_n$ be the last vertex in the construction starting from $f$, and $v_1, v_2, v_3$ be three of the vertices in $L_1$. Clearly $v_n$ is not in $L_0$ or $L_1$. We contract all edges not adjacent to any of $\{v_0, v_0', v_1, v_2, v_3, v_n\}$ (see Figure 11(a)). Since $G$ is 1-path, $v_n$ is adjacent to all of $L_1$ vertices $v_1, v_2, v_3$ in the contracted graph. The contracted graph is a $K_{3,3}$ with $v_0$, $v_0'$ and $v_n$ in one partition and $v_1$, $v_2$ and $v_3$ in the other.

**1.b** Let $v_n$ be the last vertex in the construction starting from $f$ as above, and $v_1, v_2$ be the two vertices in $L_1$. In order to be triangle-free, the third step vertex $v_3$ must be constructed with $v_1$ and $v_2$ as base pair of vertices (see Figure 11(b)), and $v_3 \neq v_n$ since $deg(v_0) \geq 3$. Contract all edges not adjacent to any of $\{v_0, v_0', v_1, v_2, v_3, v_n\}$. Since $G$ is 1-path and $deg(v_0) \geq 3$, $deg(v_0') \geq 3$, $v_n$ must be adjacent to $v_0, v_0', v_3$ in the contracted graph. The contracted graph has a $K_{3,3}$ minor with $v_0$, $v_0'$, $v_3$ in one partition and $v_1$, $v_2$, $v_n$ in the other.

**2.a** Directly follows from Lemma 5.6(1).

**2.b** Since $G$ is triangle-free, $deg(v_1) \geq 3$ and $deg(v_2) \geq 3$ is equivalent to $v_0 \notin$ the last level $L_t$ and $v_0' \notin$ the last level $L_t$, and the conclusion directly follows from 5.6(2). □

**Lemma 5.8.** *Given a 1-path, triangle-free, 1-dof tree-decomposable graph $G$ with base non-edge $f = (v_0, v_0')$, with $v_1$, $v_2$ being the only vertices in $L_1$,*

*(1) If $v_0$ is in the last level $L_t$ and $v_0'$ is not, then $G\setminus\{v_0\}$ is a 1-path, triangle-free, 1-dof tree-decomposable graph, with a 1-path construction from base non-edge $(v_1, v_2)$.*

(2) If both $v_0$ and $v_0'$ are in the last level $L_t$, then $G \setminus \{v_0, v_0'\}$ is a 1-path, triangle-free, 1-dof tree-decomposable graph, with a 1-path construction from base non-edge $(v_1, v_2)$.

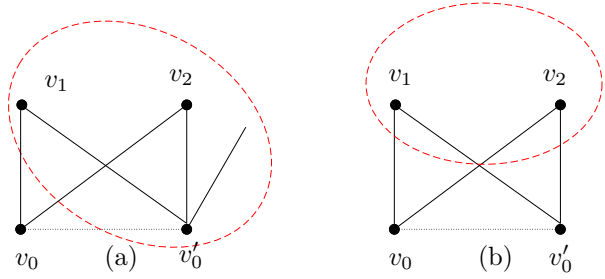

Figure 12. For proof of Lemma 5.8.

**Proof.** *1.* See Figure 12 (a). Clearly we can construct $G$ from $(v_1, v_2)$ by taking $v_0 \triangleleft (v_1, v_2)$, $v_0' \triangleleft (v_1, v_2)$ as the first two construction steps. Since $v_0$ is in the last level $L_t$ of $G$, $G \setminus \{v_0\}$ is still 1-dof tree-decomposable with base non-edge $(v_1, v_2)$. Since $G$ is triangle-free, $G \setminus \{v_0\}$ is also triangle-free. Also notice that by excluding $v_0 \triangleleft (v_1, v_2)$ from $G$, only $v_1$ and $v_2$ can have their degree decreased, so no vertices other than $v_1, v_2$ can be can be added to $L_t$ of $G \setminus \{v_0\}$, and $G \setminus \{v_0\}$ is still 1-path from $(v_1, v_2)$.
*2.* See Figure 12 (b). The argument is similar to the above. □

In fact, a similar lemma holds even if the graph is not triangle-free. It is used in Part I for proving the base case of Theorem 4.3. We state it here since it uses Lemma 5.6 above.

**Lemma 5.9.** *Let $G$ be a non-trivial 1-path, 1-dof tree-decomposable graph with base non-edge $f = (v_0, v_0')$ (by non-trivial we mean that $G$ has more than two construction steps from $f$). $G$ has low Cayley complexity on $(v_0, v_0')$ if and only if the following conditions hold:*
1. *There are only two vertices in $L_1$, $v_1$ and $v_2$.*
2. *(a) If both $v_0$ and $v_0'$ are in the last level $L_t$, $G \setminus \{v_0, v_0'\}$ is a 1-dof tree-decomposable graph with low Cayley complexity and has 1-path construction from base non-edge $(v_1, v_2)$.*
   *(b) If only $v_0$ is in the last level $L_t$, let the next construction step be $v_3 \triangleleft (u_1 \in T_5, u_1' \in T_6)$. The graph $G \setminus \{v_0\}$ is a 1-dof tree-decomposable graph with low Cayley complexity, and has a 1-path construction from base non-edge $(u_1, u_1')$ or $(v_0', v_3)$.*

**Proof.**
($\Longleftarrow$): suppose $G$ satisfies *1* and *2*. By Theorem 3.1, $G$ has low Cayley complexity on $(v_1, v_2)$.
($\Longrightarrow$): let $m$ be the number of vertices in $L_1$. Since $G$ is non-trivial, $m \geq 2$. By Lemma 5.6 (1), $m < 3$. So $m = 2$, thus Condition *1* is satisfied. By Lemma 5.6 (2), at least one of $v_1$ or $v_2$ is in the last level $L_t$. The proof of Condition *2* is similar to the proof of Lemma 5.8, except that for *(b)*, since $T_5$ and $T_6$ may have their number of shared vertices decreased, we may have a 1-path construction from either $(u_1, u_1')$ or $(v_0', v_3)$. See Figure 13. □

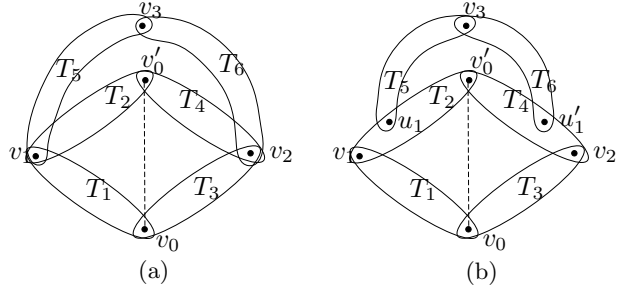

Figure 13. For proof of Lemma 5.9.

### 5.2. Limits of finite forbidden minor characterization for low Cayley complexity

**Observation 5.10.** For any graph $G_s$, there exists a 1-path 1-dof tree-decomposable graph $G$ such that $G$ has low Cayley complexity and $G_s$ is a minor of $G$.

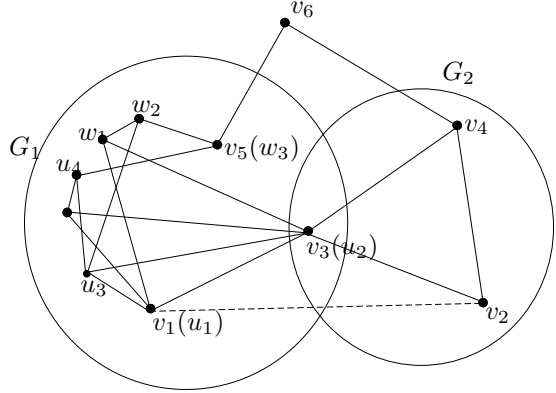

Figure 14. For Observation 5.10. A 1-path graph with low Cayley complexity on $f = (v_1, v_2)$, and has a $K_5$ minor in the cluster $G_1$. In general, it can have an arbitrary clique as a minor.

**Proof.** We prove that $G$ can have an arbitrary clique $K_m$ as minor.

Refer to Figure 14. The graph $G$ is a 1-path 1-dof tree-decomposable graph with base non-edge $f = (v_1, v_2)$ and two construction steps $v_3 \triangleleft (v_1 \in G_1, v_2 \in G_2)$, $v_6 \triangleleft (v_5, v_4)$, and it clearly has low Cayley complexity. We prove by induction on $m$, that for any $K_m$, we can construct a tree-decomposable subgraph $G_1^m$ with $K_m$ as minor and only one vertex in the last level $L_t$. The base cases ($m = 1, 2, 3$) are trivial. As the induction hypothesis, we assume that we can construct $G_1^m$, and show that we can construct $G_1^{m+1}$ from $G_1^m$. We pick $m$ vertices $u_1, \cdots, u_m$ from $G_1^m$, which correspond to the $m$ vertices of the minor $K_m$, where $u_m$ is the only vertex in the last level of $G_1^m$. We add construction steps: $w_1 \triangleleft (u_1, u_2)$, $w_2 \triangleleft (w_1, u_3)$, $w_3 \triangleleft (w_2, u_4)$, till $w_{m-1} \triangleleft (w_{m-2}, u_m)$ to get $G_1^{m+1}$ (refer to Figure 14 for a $K_5$ example). Clearly, $G_1^{m+1}$ is still a tree-decomposable graph with $w_{m+1}$ as the only vertex in the last level $L_t$. By contracting all the edges adjacent to any vertex other than $u_1, \cdots, u_m$ and $w_{m-1}$, we get a $K_{m+1}$ minor. By connecting $w_{m-1}$ to $v_6$, the 1-path property of $G$ is maintained. Thus, we have proved that $G$ can contain a $K_{m+1}$ minor. □

**Observation 5.11.** For any graph $G_s$, there is a triangle-free 1-dof tree-decomposable graph $G$ such that $G$ has low Cayley complexity and $G_s$ is a minor of $G$.

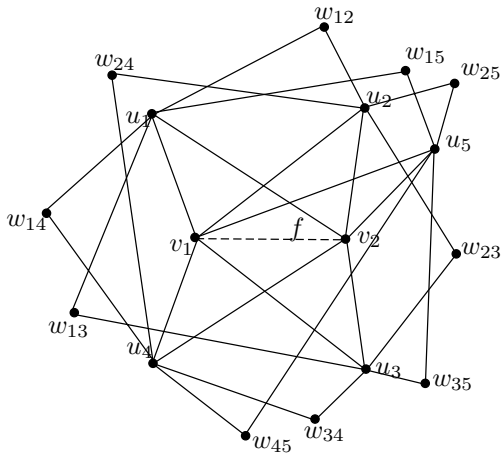

Figure 15. For Observation 5.11. A triangle-free graph with low Cayley complexity on $f = (v_1, v_2)$, and has a $K_5$ minor. In general, it can have an arbitrary clique as a minor.

**Proof.** We prove that $G$ can have an arbitrary clique $K_m$ as minor.

See Figure 15. Starting from base non-edge $f = (v_1, v_2)$, we first construct $m$ vertices in $L_1$: $u_i \triangleleft (v_1, v_2)$, where $1 \leq i \leq m$. Then we construct $\binom{m}{2}$ vertices in $L_2$ such that $w_{ij} \triangleleft (u_i, u_j)$ for all $i \neq j$. Clearly $G$ is 1-dof triangle-free with low Cayley complexity. Each vertex $u_i$ in $L_1$ corresponds to a vertex of $K_m$. □

## 6. Conclusion

In Part II of this paper, we proved that low Cayley complexity is not affected by the choice of base non-edge, and is thus a property of the graph. We also gave an efficient algorithm to verify whether a graph has low Cayley complexity. Implementation of this algorithm is part of our new software CayMos (under development), web-accessible at http://www.cise.ufl.edu/˜menghan/caymos.html. A different manuscript (Sitharam and Wang, 2012) describes CayMos together with an analysis of common and well-known mechanisms, as well as comparison with other softwares with related functionalities.

In addition, we proved that a finite forbidden-minor characterization of low Cayley complexity only exists for 1-path, tree-decomposable, 1-dof tree-decomposable graphs, and for such graphs low Cayley complexity is equivalent to graph planarity.